# First-Principle Investigation Of Near-Field Energy Transfer Between Localized Quantum Emitters in Solids

Swarnabha Chattaraj,[1*] Supratik Guha,[1,2] Giulia Galli[1,2*]

*1. Materials Science Division, Argonne National Laboratory*

*2. Pritzker School of Molecular Engineering, University of Chicago*

*Emails: gagalli@uchicago.edu, schattaraj@anl.gov*

**Abstract**

We present a predictive and general approach to investigate near-field energy transfer processes between localized defects in semiconductors, which couples first principle electronic structure calculations and a nonrelativistic quantum electrodynamics description of photons in the weak-coupling regime. We apply our approach to investigate an exemplar point defect in an oxide, the F center in MgO, and we show that the energy transfer from a magnetic source, e.g., a rare earth impurity, to the vacancy can lead to spin non conserving long-lived excitation that are dominant processes in the near field, at distances relevant to the design of photonic devices and ultra-high dense memories. We also define a descriptor for coherent energy transfer to predict geometrical configurations of emitters to enable long-lived excitations, that are useful to design optical memories in semiconductor and insulators.

# Introduction

Energy transfer between localized emitters embedded in a solid host material, for example point-defects, is a ubiquitous phenomenon of interest to several fields, including photonics, microelectronics, and quantum networks. Energy transfer may be mediated by several, complex mechanisms, including direct tunnelling, photons, and phonons. Here we focus on photon mediated transfer from source to absorber- i.e., Fluorescent Resonance Energy Transfer (FRET) or Non-Radiative Energy Transfer (NRET) [1-5], and we consider the near field regime where the source-absorber distance is smaller than the wavelength of the photons being transferred.

This regime is of interest, for example, to study quantum emitters in close proximity of each other in semiconductors and insulators, and/or emitters close to other impurities such as oxygen vacancies in oxide hosts. Understanding NRET phenomena in these systems is critical to gain insight into the design of solid-state rare earth doped quantum memories and repeaters [6], as well as, potentially, ultra-high density classical optical memories. In particular, we envision to individually address, by optical means, narrow band rare earth (RE) emitters [7] out of an ensemble dispersed in a solid-state host, for example an oxide, and to transfer narrow-band excitations to a proximal defect (see Fig.1). In such platforms, with a realistic few ppm doping, the average separation between REs and defects can be of the order of ~5 to ~10 nm—a distance much smaller than the wavelength of the optical/near IR photons (~500nm to ~$1\mu m$). Thus, NRET processes at the near field regime play a critical role. We also envision enhancing the lifetime of the excitation transferred between the RE and a nearby defect through spin-non conserving transitions.

In the near field, transitions mostly occur via virtual photons [8] resulting in absorption and emission processes with characteristics markedly different from those of electronic transitions occurring in the far field regime and may result in violation of spin and orbital selection rules [8-11] that are valid in the far-field. The ability to engineer spin- or orbital-forbidden optical absorption excitations in emitters at near field provides a pathway to significantly increase the lifetime of optical memories [Supplementary Information (SI)]. The material systems relevant to such technologies span in principle a wide range including native defects [12, 13] (F centers, Frenkel defects), implanted defects, and deposited dopants (rare earth in oxides [6,7, 14-19], NV and SiV in diamond [20-23], donor or acceptors in semiconductors), and quantum dots [24-26].

Investigating such vast range of material systems demands a theoretical approach that allows for an efficient way to accounting for the symmetry, spread, and many-body nature of the wavefunctions of the localized emitters and hence it requires the use of first-principle electronic structure theories such as density functional theory (DFT) [27-29], many body perturbation theories [30-32] or, if multi-reference states are present, more sophisticated approaches such as quantum embedding [33-36] theories and time dependent DFT [37].

Historically the study of resonant energy transfer has developed in the field of molecular quantum electrodynamics (QED) to address energy transfer between molecules and molecular complexes [38-44]. In the solid state, NRET processes have also been explored to understand exciton diffusion, transfer, and light trapping mechanisms [3, 45]. In either case, dipole-dipole approximations were often adopted, and the matrix elements for photon absorption and emission defined phenomenologically.

On the other hand, the approach of quantum electrodynamical density functional theory (QEDFT) [46-49] treats the light matter coupling entirely from first principles with a full diagonalization of the electron-photon coupled system (polaritons and exciton-polaritons). Such an approach is appropriate to describe a strong-coupling regime where the light-matter coupling can be comparable to the energy of the photon itself. While QEDFT enabled unique applications in polaritonic chemistry, for example polaritonic catalysis [48-49], the approach is computationally rather demanding, and it typically considers a single cavity mode coupled with localized electronic systems such as isolated molecules.

However, there are several instances of weak coupling between light and matter, in particular quantum emitters in solids, where a full polariton treatment is neither necessary nor feasible, and a perturbative approach is warranted. Here we model the light matter coupling at localized emitters in the solid state from first principles. We use the Pauli Hamiltonian [50-53] to describe the light-matter coupling with the photon field operators expressed in a multipole basis [54-56]. The properties of the multipole basis are exploited to increase the efficiency of our calculations, paving the way to tackle large ensembles of defects which would be difficult to treat in a plane wave basis. As an example, we use our approach to investigate energy transfer in a realistic system of localized defects in a typical insulating host, MgO with F centers, which are very commonly found in real substrates. We consider a magnetic and an electric dipolar source emitting at near field (mimicking a rare earth substitutional site in the host), and we investigate NRET processes as a function of the source-to-absorber distance and orientation. We also discuss configurations and criteria to obtain dominant spin-flip transitions in the near field, which are relevant for the design of classical optical memories and of quantum networks.

In the rest of the paper, we lay out the theoretical approach that integrates first principle electronic structures, and non-relativistic QED to study NRET between arbitrary localized emitters in solids from nanometer to macroscopic separations. We then apply our approach to the F center in MgO and highlight the key differences between near field and far field absorption processes, and we define an effective parameter for coherent transport rates and discuss its implication to device design.

## Results

We start by presenting our theoretical approach to calculate the time dependent probability amplitude for the NRET between two emitters at an arbitrary separation and orientation in a solid host, and we then present our results for an oxygen vacancy in MgO.

**Theoretical approach**

We assume that the main transitions of interest to NRET between quantum emitters are those between the localized bound states of electrons within the source and the absorber (Fig. 1), and do not consider any transitions between the bulk extended states [57]. We further assume that the emitters are sufficiently distant from each other so that there is no overlap between their respective electronic wavefunctions. We write the Hamiltonian of the coupled emitter-photon system as:

$$H = H_S + H_A + H_{Field} + H_{int} \qquad (1)$$

Here $H_S$ and $H_A$ are the Hamiltonians of the source S (for example a rare earth impurity in an oxide) and absorber A (for example an oxygen vacancy in an oxide) respectively, and $H_{Field}$ is the Hamiltonian of the photon. We define $H_0 = H_S + H_A + H_{Field}$ as the unperturbed Hamiltonian and $H_{int}$ as the light-matter interaction perturbation on $H_0$.

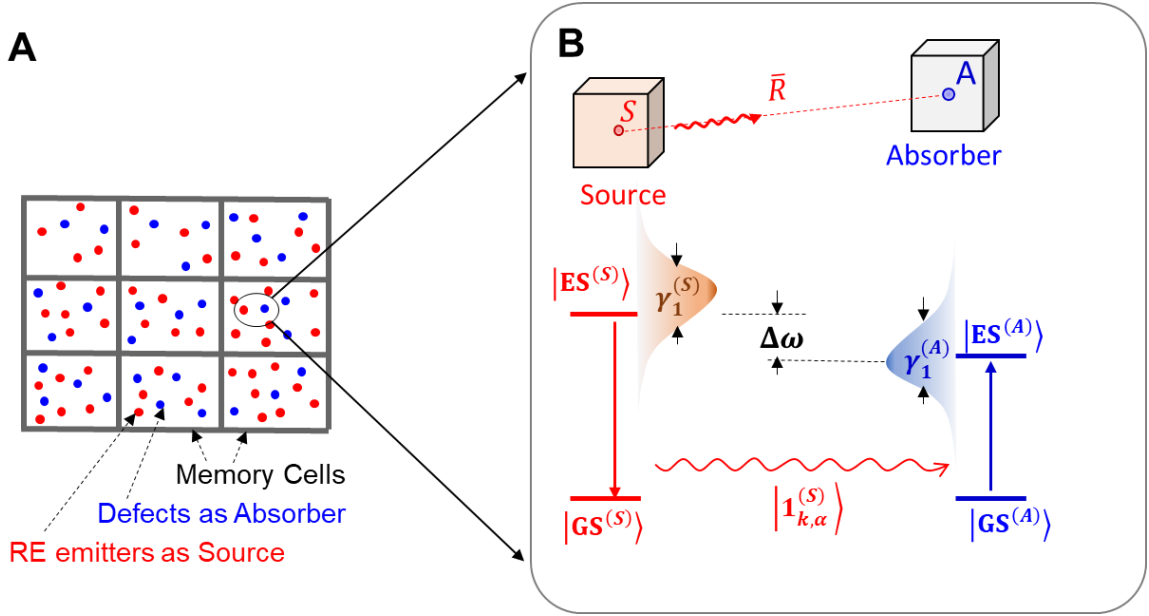

**Fig. 1. (A)** A schematic representation of ultra-high density optical memories (see SI) where each memory-cell in a solid host contains an ensemble of rare-earth (RE) emitters (source, red) and point-defects (absorber, blue). **(B)** A RE emitter-defect complex representing the working "unit" of the memory. Optical excitations of RE emitters that are spectrally separated can be transferred to a suitable defect in the proximity of the emitter to trap the excitation and increase its lifetime. The source (S) and the absorber (A) are separated by $\bar{R}$. The photon propagation (curly arrow) is treated within quantum electrodynamics. The ground and excited states of S and A are labeled with GS and ES, respectively and the photon by $|1_{k,\alpha}\rangle$, where $k$ indicates the photon momentum and $\alpha$ represents the set {L = orbital angular momentum, Jz = total angular momentum projected to a chosen direction z, P = parity} specifying the multipole mode of the photon. The linewidth of the transition at finite timescale is $\gamma_1^{(S/A)} = 1/T_1^{(S/A)}$ ($T_1$ is the decay lifetime); due to the finite linewidth, energy transfer may occur even if energy levels are mismatched by a frequency difference $\Delta\omega$.

The many electron eigenstates $|\Phi^{(S/A)}\rangle$ of the Hamiltonians $H_{S/A}$ can be written as a sum of Slater determinants constructed from single electronic states $\{|\phi_i^{(S/A)}\rangle, \ i = 1:N_{S/A}\}$ where $N_{S/A}$ is the number of the single particle electronic states in the source $S$ or the absorber A. The field Hamiltonian is expressed as $H_{Field} = \sum_{k,\alpha} \hbar\omega_k \left(a_{k,\alpha}^\dagger a_{k,\alpha} + \frac{1}{2}\right)$, where $a_{k,\alpha}^\dagger$ is the creation operator of a photon in the mode $\{k,\alpha\}$; $k$ is the radial wave number and $\alpha$ represents the set $\{L, J_Z, P\}$ specifying the symmetry and polarization— $L$ being the photon's orbital angular momentum, $J_Z$ the photon's total angular momentum (orbital and spin) projected along a chosen direction z, and $P$ the parity. The state of the photon, emitted from the source $S$, is represented as a linear combination of radiating multipole modes [54-56, SI] denoted by $\left|1_{k,\alpha}^{(S)}\right\rangle$. The superscript S denotes that the multipole modes are centered at the source S. These modes, obtained by solving the Maxwell equations, are standing waves, with a Bessel j function as radial part, for a perfectly bound spherical cavity. For an open cavity the radial functions are instead type 1 Hankel functions [see SI for further details].

We write $H_{int}$ as:

$$H_{int} = \sum_{E=S,A} \sum_{i=1}^{N_E} \left[ -\frac{e\,\overline{p}_i \cdot \overline{A}}{2m_0} - \frac{e\,\overline{A}\cdot\overline{p}_i}{2m_0} + eA_0 + g\frac{e\hbar}{2m_0}\overline{\sigma}_i \cdot \overline{\nabla} \times \overline{A} \right]. \quad (2)$$

Here, $\overline{p}_i$ and $\overline{\sigma}_i$ are the momentum operator and the Pauli spin operator of the $i$th electron, $m_0$ the rest mass of electrons, $e$ the electron charge, $\hbar$ the Plank constant; g is the Lande g-factor of the electron which is equal to 2 for free electrons and may take different values due to magnetic screening. The g-factor, directly probed in electron paramagnetic resonance (EPR) measurements, can be computed using DFT [58]. The operators $\overline{A}$ and $A_0$ are the magnetic vector potential and scalar potential field of the photon. Their action on a multipole state $\left|1_{k,\alpha}^{(S)}\right\rangle$ with $\alpha = \{L, J_z, P\}$ is expressed as: $\overline{A}\left|1_{k,\alpha}^{(S)}\right\rangle = \overline{A}_{k,\alpha}^{(S)}(\bar{r},t)|0\rangle$ and $A_0\left|1_{k,\alpha}^{(S)}\right\rangle = A_{0;\,k,\alpha}^{(S)}(\bar{r},t)|0\rangle$. Here $\bar{r}$ is the position in the crystal referred to the source, and the fields $\overline{A}_{k,\alpha}^{(S)}(\bar{r},t)$ and $A_{0;\,k,\alpha}^{(S)}(\bar{r},t)$ can be expressed analytically. For an electric type of mode $((-1)^L = P)$:

$$\overline{A}_{k,\alpha}^{(S)}(\bar{r},t) = \frac{1}{4\pi}\sqrt{\frac{k}{R_{norm}}} \left[ \left( \sqrt{\frac{L}{2L+1}}\, g_{L+1}(kr)\, \overline{Y}_{L,L+1,J_z}(\hat{r}) + \sqrt{\frac{L+1}{2L+1}}\, g_{L-1}(kr)\, \overline{Y}_{L,L-1,J_z}(\hat{r}) \right) \right.$$

$$+ C\left( -\sqrt{\frac{L+1}{2L+1}}\, g_{L+1}(kr)\, \overline{Y}_{L,L+1,J_z}(\hat{r}) \right.$$

$$\left.\left. + \sqrt{\frac{L}{2L+1}}\, g_{L-1}(kr)\, \overline{Y}_{L,L-1,J_z}(\hat{r}) \right) \right] e^{-i\omega_k t} \quad (3)$$

and

$$A^{(S)}_{0;k,\alpha}(\bar{r},t) = \frac{C}{4\pi}\sqrt{\frac{k}{R_{norm}}}\, g_L(kr)\, Y_{L,J_z}(\hat{r})e^{-i\omega_k t}. \tag{4}$$

Here $g_L(kr) = 4\pi i^L z_L(kr)$, $z_L(kr)$ are spherical Bessel functions, $Y_{L,M}$ and $\bar{Y}_{J,L,M}$ are spherical scalar and vector harmonics respectively, and $R_{norm}$ denotes the radius of the normalizing sphere for the multipole mode (see SI for details). For a magnetic type of mode ($(-1)^L = -P$), the vector and scalar potential fields are:

$$\bar{A}^{(S)}_{k,\alpha}(\bar{r},t) = \frac{1}{4\pi}\sqrt{\frac{k}{R}}\, g_L(kr)\, \bar{Y}_{L,L,J_z}(\hat{r})e^{-i\omega_k t} \tag{5}$$

and

$$A^{(S)}_{0;k,\alpha}(\bar{r},t) = 0 \tag{6}$$

The constant C is an arbitrary parameter representing gauge freedom. The choice of C does not affect the value of the matrix elements discussed below, but the computation can be simplified by choosing $C = 0$ (radiation gauge) for which the scalar potential vanishes (Eq. (4)). Eq. (2), derived from the generalized Kramer-Heisenberg or Pauli Hamiltonian [50] reduces to $e\bar{E}\cdot\bar{r}$ only under specific conditions [53] -- (1) resonant conditions i.e. when the frequency of the photon matches the energy gap of the electronic states and when (2) the electronic states are eigenfunctions of a *local* Hamiltonian. The first criterion does not apply for NRET where significant energy transfer may occur under non-resonant conditions. In addition, in DFT calculations with hybrid functionals, non-local terms are introduced in the Hamiltonian, in addition to non-local terms present in the pseudopotentials. Thus the fully general $\bar{A}\cdot\bar{p}$ form (Eq.(2)) should be used when first principle hybrid DFT calculations are carried out.

As mentioned in the introduction, in non-relativistic QEDFT [46-48] the Hamiltonian (Eq.(1)) is diagonalized for arbitrary light matter coupling strengths and coupled electrons and photons (polaritons and exciton-polaritons) wavefunctions are computed. Such an approach allows for the exploration of strong coupling regimes useful to investigating phenomena such as polaritonic catalysis [49]. However, in the case of NRET between localized defects in a bulk solid, the photon modes leak into the continuum within ~femtosecond timescale while light matter coupling (as measured by radiative decays) usually occurs within ~nanosecond or longer time scales for typical emitters, including deep levels and rare earths in semiconductors and insulators, and quantum dots. These conditions define the so-called weak coupling regime- whose description may be obtained using a perturbative approach, where the unperturbed states of electrons and photons are used as basis sets for the interacting Hamiltonian of Eq. (1).

We consider an initial state where only the source S is in an excited state (ES), no photon is present in the system and the absorber is in the ground state (GS): $\Psi_i = \Psi(t=0) = |ES^{(S)}\rangle|GS^{(A)}\rangle$; after the energy is transferred from S to the absorber A, the system is in the final state $\Psi_f = |GS^{(S)}\rangle|ES^{(A)}\rangle$. The energy transfer between S and A can occur through all possible intermediate states $|I\rangle = |GS^{(S)}\rangle|GS^{(A)}\rangle|1_{k_A,\alpha}^{(S)}\rangle$.

The NRET amplitude $c(t) = \langle \Psi_f | \Psi(t) \rangle$ is [4,5]:

$$c(t) = \frac{2M}{\hbar} \frac{\sin\left(\frac{\Delta\omega}{2} t\right)}{\Delta\omega} e^{-\frac{t}{2T_{1S}}} \qquad (7)$$

where $\Delta\omega = \omega^{(A)} - \omega^{(S)}$ is the energy mismatch between the source and the absorber as indicated in Fig. 1, $T_{1S}$ is the decay rate of the isolated source, and within second order perturbation theory, the matrix element $M$ is expressed as a sum over all possible intermediate states:

$$M = \sum_I \frac{\langle \Psi_f | \widetilde{H}_{int} | I \rangle \langle I | \widetilde{H}_{int} | \Psi_i \rangle}{\hbar(\omega_i - \omega_I)} \qquad (8)$$

where $\widetilde{H}_{int}(t) = e^{\frac{iH_0 t}{\hbar}} H_{int}(t) e^{-\frac{iH_0 t}{\hbar}}$. For the absorption of a photon in the state $|1_{k,\alpha}^{(S)}\rangle$ resulting in an electronic transition from $|GS^{(A)}\rangle$ to $|ES^{(A)}\rangle$ at the absorber site, we define the photon absorption matrix element:

$$V_{k,\alpha}^{(A)} = \langle \Psi_f | \widetilde{H}_{int} | I \rangle = \langle ES^{(A)} | \widetilde{H}_{int} | GS^{(A)}, 1_{k,\alpha}^{(S)} \rangle. \qquad (9)$$

Eq. (9) can be simplified to an expression containing only integrals of single electron orbitals $\phi_i^{(B)}$ using the Slater-Condon rule [59, 60] (see SI):

$$V_{k,\alpha}^{(A)} = \left\langle \phi_i^{(A)} \left| \widetilde{H}_{int} \right| \phi_j^{(A)}, 1_{k,\alpha}^{(S)} \right\rangle \tag{10}$$

Similarly, for the emission of a photon $|1_{k,\alpha}^{(S)}\rangle$ resulting from an electronic transition from the excited state $|ES^{(S)}\rangle$ to the ground state $|GS^{(S)}\rangle$, we define the matrix element:

$$V_{k,\alpha}^{(S)} = \langle I|H_{int}|\Psi_i\rangle = \left\langle \phi_i^{(S)}, 1_{k,\alpha}^{(S)} \left| \widetilde{H}_{int} \right| \phi_j^{(S)} \right\rangle. \tag{11}$$

As discussed in the SI, it is convenient to use the following expressions for the matrix elements $V_{k,\alpha}^{(S)} = v_\alpha^{(S)}(k)\sqrt{\Delta k}$, and $V_{k,\alpha}^{(A)} = v_\alpha^{(A)}(k)\sqrt{\Delta k}$. Here $\Delta k$ is the width of the mode $|1_{k,\alpha}^{(S)}\rangle$ in k-space and is related to the radius of the normalizing sphere $R$: $\Delta k = \pi/R_{norm}$. From equation (8) we have:

$$M = \frac{i\pi n_i}{\hbar c} \sum_\alpha v_{k_S,\alpha}^{(S)} \, v_{k_S,\alpha}^{(S)} \tag{12}$$

where the expression of $M$ is general and contains all the multipolar contributions along with the absorption and emission matrix elements defined from first principles. Here $n_i$ represents the refractive index of the host material and $c$ the vacuum speed of light. Under the dipole approximation the expression in Eq. (12) is consistent with $M(\bar{R}) = \bar{p}_S \cdot \bar{\bar{G}}(\bar{R}) \cdot \bar{p}_A$, where $\bar{\bar{G}}$ is the electromagnetic Green function, and $\bar{p}_S$ and $\bar{p}_A$ are the transition dipoles at the source and the absorber. For convenience we can express the transfer amplitude $c(t)$ as a function of the mediating photon amplitude $c_k(t)$, with $c(t) = \int dk c_k(t)$, where (details in SI):

$$c_k(t) = \frac{1}{\hbar^2} \sum_\alpha \frac{v_{k\alpha}^{(S)} v_{k\alpha}^{(A)}}{\omega_k - \omega_S} \left( \frac{e^{-i\omega_S t} - e^{-i\omega_A t}}{\omega_A - \omega_S} - \frac{e^{-i\omega_k t} - e^{-i\omega_A t}}{\omega_A - \omega_k} \right) \tag{13}$$

Equation (13) is analogous to similar expressions using a plane wave basis (for example Eq.(1) of Ref [4]). However, here the momenta of the multipole modes are expressed by a number ($k$ =radial momentum) and not a vector (as in plane wave basis), thus simplifying the expressions particularly for the higher order multipoles [10, 11] of interest to the discussion of our results.

We note that expressing $v_{k_S,\alpha}^{(A)}$ and $v_{k_S,\alpha}^{(S)}$ from first principle allows for the study of different, interesting cases of emitters in solids: (i) emitters with electronic states localized in <nm scale, which can be described using DFT [27-29], many body perturbation theories (MBPT) [30-34], or embedding theories [35-37]; (ii) emitters such as rare earth substitutional sites where one can derive the associated multipoles of the transitions either from theory, including crystal field theory [61-63], or from experiments [6,7]; and (iii) emitters such as quantum dots for which the wavefunctions of the bound electron and hole may extend over ~20 nm and may be described using a single particle picture, together with our perturbative approach with higher order multipoles [64]. In addition, first-principle calculations of the matrix elements pave the way to predictive discoveries of viable defects and host platforms for optical memories and quantum networks.

As an example, we now turn to describe energy transfer in a well-known defect in MgO, the F center, where we assume photon sources are provided by rare-earth defects implanted in the material and pinpoint the uniqueness and the relevance of the near field effects.

**F center in MgO**

In recent years MgO has been studied as a potential host for spin defects to build quantum memories and quantum networks [65]. In such systems, it is common to observe energy transfer between spin defects acting as quantum emitters and oxygen vacancies present in the surrounding environment, resulting in dephasing of the spin defects. Similarly, in substrates with coexisting rare-earth dopants and oxygen vacancies, NRET from rare earths to the localized states of the oxygen vacancies constitute a possible excitation process. Particularly if such excitation process is dominantly a spin-flip transition, it could be used to create long lived defect excitations.

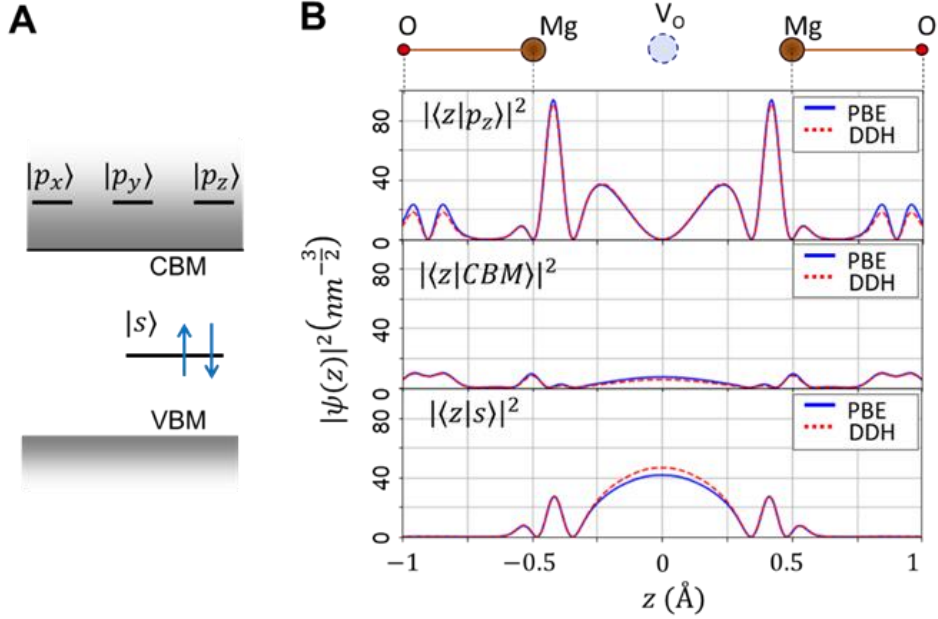

**Fig. 2. (A)** Schematic representation of the energies of the $|s\rangle$ and $|p\rangle$ states localized at the $V_O$: MgO that participate in optical absorption and emission transitions. The conduction band minimum and valence band maximum of the host are indicated by CBM and VBM, respectively. **(B)** Projections of the square moduli of the Kohn Sham wavefunctions $|\psi(z)|^2$ of the $|s\rangle$ and $|p_z\rangle$ states along a chosen direction z, computed with semi local (PBE) and hybrid (DDH) functionals [details in SI]. Owing to the highly localized nature of the orbitals, the inclusion of the exact exchange in the KS Hamiltonian has little effect on the orbital, as seen by the difference between PBE and DDH results.

Here we consider the energy transfer to an F center, i.e., a neutral oxygen vacancy ($V_O$: MgO), which has two localized electrons. The $V_O$: MgO has been experimentally and theoretically studied for several decades using optical absorption, photoluminescence, and electron spin resonance techniques [66-68]. Experimentally, the optical absorption of neutral F centers is found at ~5eV and emission at ~2.3eV and ~3eV [67,68]. As indicated in Figure 2(A) and (B), the relevant single electron orbitals are the localized s-type ($a_{1g}$) orbitals that are mid-gap, filled orbitals in the ground state, and the empty triply degenerate localized p-type ($t_{1u}$) orbitals (referred to as $p_{x, y, z}$) above the conduction band minimum (CBM). An excitation from the many-body ground $^1A_{1g}$ state may result in either excited singlet ($^1T_{1u}$) or triplet ($^3T_{1u}$) states. The ~5eV absorption is assigned to the singlet-to-singlet transition $^1A_{1g} \rightarrow {}^1T_{1u}$ [35-37, 67, 68, 70-73] and the singlet-to-triplet absorptions are known to be forbidden at far field. There has been much debate regarding the nature of the emission [66-68, 70-73]. Recent work has shown that the ~2.3 eV emission likely originates from the transition from the triplet $^3T_{1u}$ to the ground state $^1A_{1g}$ [36, 37], whereas the ~3eV emission is likely from a bound exciton [37]. All studies of optical absorption and emission on the V$_O$: MgO reported so far have only addressed the far field regime. Below we explore the near field regime.

**Transition matrix elements as a function of distance at near field**

When the distance between the emitters in a host is small compared to the wavelength of the photon being exchanged, the lifetime of the photon is shorter than its time period. The energy of such a short-lived photon (commonly referred to as virtual photon [8]) exhibits a large quantum uncertainty, resulting in an electric and magnetic field that vary as ~1/R$^3$ [1] – R being the distance from the source (see Fig.1). As R increases, the photon's propagation characteristic and its lifetime vary [9], eventually leading to an ~1/R dependence of the E and H fields at larger distances. In this regime, the spatial extent of the emitter and the absorber are usually ignored relative to the wavelength of the photon and the dipole approximation is adopted to describe the photon wavefunction. Instead, at the near field, the energy density and the E- and H-fields of the photon may vary significantly over the spatial extent of an absorber, and higher order multipoles cannot be neglected when describing the energy transfer processes [2, 11].

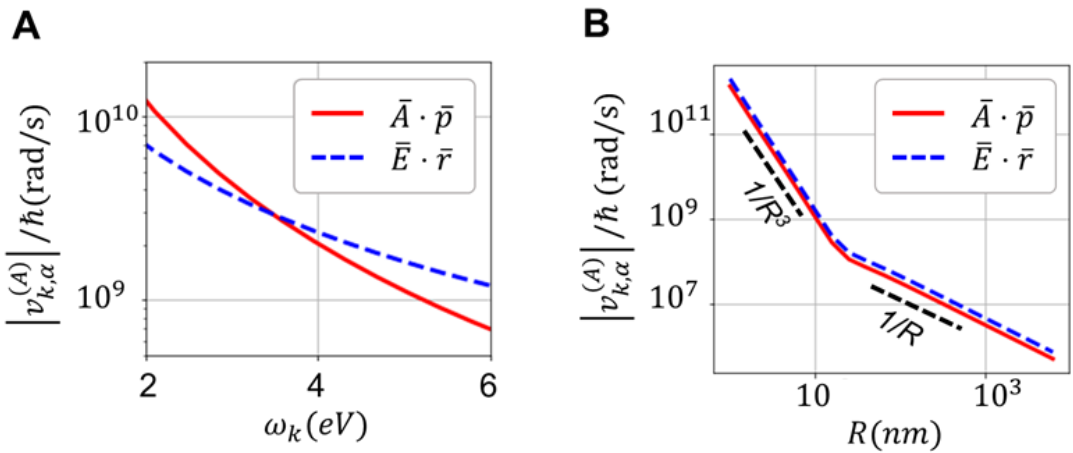

**Fig. 3.** Comparison between the absorption matrix elements for the ground to excited state singlet transition for V$_O$: MgO ($v_{k\alpha}^{(A)}$), computed using the $A \cdot p$ Hamiltonian (red curve, see text) and $E \cdot r$ Hamiltonian (blue curve). Panel **(A)** shows the frequency response at a constant source-to-absorber distance of 10 nm and panel **(B)** shows the distance dependence at resonance condition at 5eV.

We numerically computed the matrix elements $v_{k,\alpha}^{(A)}(\bar{R})$ (Eq. (10) and Eq. (2) with g=2) for an oxygen vacancy in MgO. The orbital wavefunctions (Fig. 2(B)), are calculated using first principle DFT calculations (see SI). We first verified that the transition between the 1/R and 1/R$^3$ regimes occurs, as expected, at R~ 1/k; for the 5eV optical absorption line this distance corresponds to ~ 20 nm. Separations of <20nm can be readily reached at realistic doping concentrations of ~10 ppm, indicating that near field energy transfer is a relevant process in devices. In Figure 3 (A)&(B) we also show the comparison between results obtained with the $\bar{A} \cdot \bar{p}$ and $\bar{E} \cdot \bar{r}$ Hamiltonians for s to p$_z$ type spin conserving transitions ($|GS\rangle$ to $|ES_s\rangle$). Note that, due to the non-local nature of the Kohn-Sham Hamiltonian, results obtained with $\bar{A} \cdot \bar{p}$ and $\bar{E} \cdot \bar{r}$ differ. (For comparison with emitter with a local Hamiltonian, see Section S5 of the SI.)

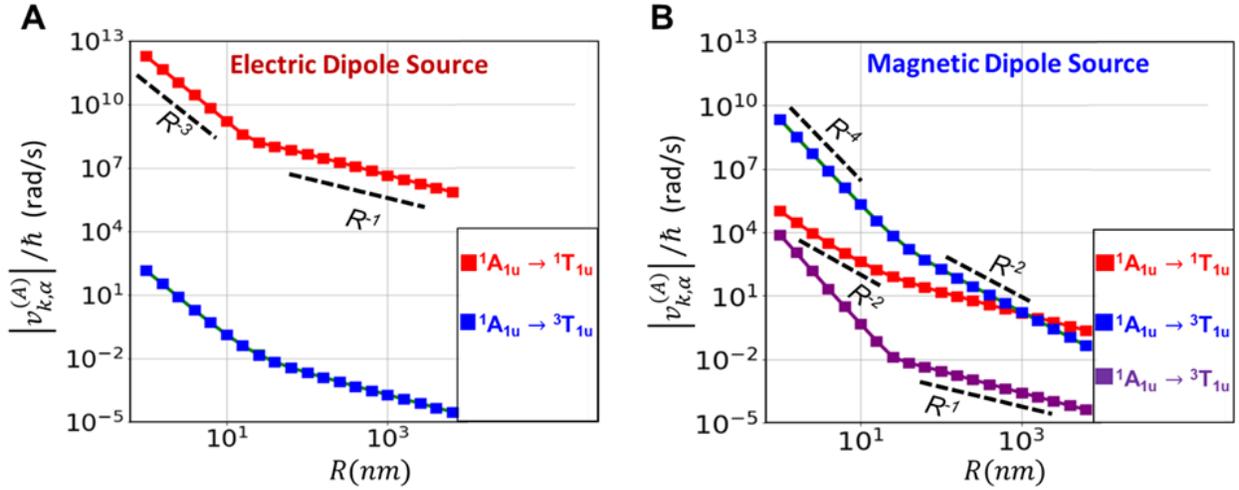

**Fig. 4.** The distance dependence (R, see Fig.1) of the matrix element $|v_{k,\alpha}^{(A)}(\bar{R})|$ at a fixed direction $\hat{R} = \hat{X}$ for the spin conserving ($^1A_{1u} \to {}^1T_{1u}$; red squares) and spin non conserving ($^1A_{1u} \to {}^3T_{1u}$; blue and purple Squares) absorption transitions in the V$_O$: MgO center for **(A)** Electric dipole source and **(B)** Magnetic dipole source. Blue and purple squares indicate the transition to $^3T_{1u}$ ($m_s = \pm 1$) and to $^3T_{1u}$ ($m_s = 0$), respectively. The power low behavior of the matrix elements as a function of R is shown for each matrix element. The results indicate that for energy transfer from a magnetic dipole source, a dominant spin-non conserving transition can be achieved in the near field.

In V$_O$: MgO, while the transitions between the GS and the triplet states of the F center are spin forbidden at far field, they have nonzero amplitude at near field, due to the selection rules of the photon emission and absorption when transitions higher than dipolar ones are involved [10].

Hence the sum of the total angular momenta of the electrons in the source and the absorber may not be conserved at near field.

We show in Figure 4(A) and (B) the matrix elements for the GS to singlet and triplet transitions where the near field source (S) is either of an Electric Dipole (ED) or Magnetic Dipole (MD) type. Note, for rare earth emitters, both ED and MD type source can naturally exist [6, 7]. The case for an ED source $(\alpha = \{L = 1, J_z = 0, P = -1\})$ is shown in Fig. 4(A), where the singlet transition is dominant both at near and far field. However, we see that the GS to $|\,^3T_{1u}(m_s = \pm 1)\rangle$ triplet transition is allowed at the near field. The $|GS\rangle$ to $|\,^3T_{1u}(m_s = 0)\rangle$ transition remains forbidden even at near field, as the symmetry between the spin up and spin down states is not broken. By including the effect of zero field splitting, or applying an external B field, one can further reduce the symmetry, in which case the $|GS\rangle$ to $|\,^3T_{1u}(m_s = 0)\rangle$ transition may become active. For example, for a photon state with preferred direction of rotation along a given axis z, the symmetry between the up and down spin is broken and the matrix element for the transition between the singlet and the $m_s = 0$ triplet state becomes non-zero.

The case of a magnetic source is shown in Fig 4(B). As in Fig. 4(A), the absorber and the source are assumed to lay in the X direction. In this case, the spin flip singlet-to-triplet transition ($|GS\rangle$ to $|\,^3T_{1u}(m_s = \pm 1)\rangle$) becomes dominant, compared to the spin-conserving singlet-to-singlet ($|GS\rangle$ to $|\,^1T_{1u}\rangle$) one. This result highlights the striking difference between far-field and the near field absorption processes. The spin-flip singlet-to-triplet transition is expected to give rise to a long-lived excited state at the F center and carries great significance towards the proposed optical memory platform (see SI), where excitations are transferred from RE emitters to nearby oxygen vacancies. Note that magnetic dipole radiation naturally exists for 4f-4f and 4f-5d transitions in several platforms of rare earth doped oxide systems [6,7]. Further, the z-directional magnetic dipole source produces a z-directional magnetic field at the absorber, leading to a symmetry breaking between the spin up and down states. This symmetry reduction results in a nonzero absorption matrix element for the ($|GS\rangle$ to $|\,^3T_{1u}(m_s = 0)\rangle$) transition as shown by the purple curve. Interestingly, at large distances, the spin conserving transition (red curve) becomes dominant again- highlighting the importance of the explicit treatment of the absorption of a photon at the near field, as a function of the distance. Overall, the results reported in Fig. 4 provide guidance and insight towards singlet-to-triplet absorption processes that may be used for the design of optical memories.

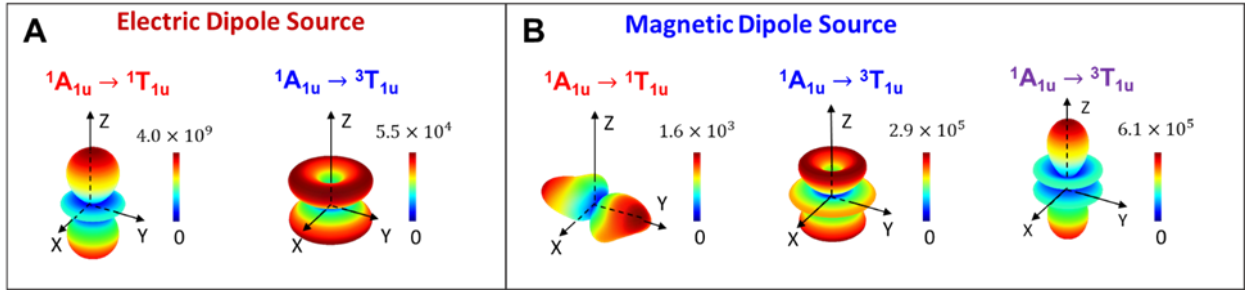

**Fig. 5.** The angular dependence of the absorption matrix element $|v_{k,\alpha}^{(A)}(\bar{R})|$ at a fixed source-absorber distance of $R=10$ nm for the spin conserving and spin non conserving absorption transitions in the $V_O$: MgO center for **(A)** Electric dipole source and **(B)** Magnetic dipole source. The color code for transitions is the same as in Fig.4. The color scale represents the value of $\left|v_{k,\alpha}^{(A)}\right|/\hbar$ in the units of rad/s.

We get further insight into the difference between far and near field processes by studying the angular distribution of the absorption matrix elements. We fix the source-absorber distance at 10 nm (near field in our case) and vary the source-to-absorber direction $\hat{R}$. For the ED and MD sources, $|v_{k,\alpha}^{(A)}(\hat{R})|$ is plotted as a function of $\hat{R}$ and the resulting angular distributions are shown in Fig. 5(A) and (B), respectively, for the spin-conserving ($|GS\rangle$ to $|{}^1T_{1u}\rangle$) and spin-non conserving ($|GS\rangle$ to $|{}^3T_{1u}(m_s = \pm 1)\rangle$) (blue), and ($|GS\rangle$ to $|{}^3T_{1u}(m_s = 0)\rangle$) (purple) transitions. Note that the angular symmetry of the spin-conserving and spin-non conserving transitions are different for ED and MD sources. This difference is interesting as it implies that at certain specific orientations between the source and the absorber, spin non conserving processes can be engineered to become dominant; hence our results provide insight into geometrical configurations which may be attained by nanofabrication design to control the angular orientation between the source and absorber.

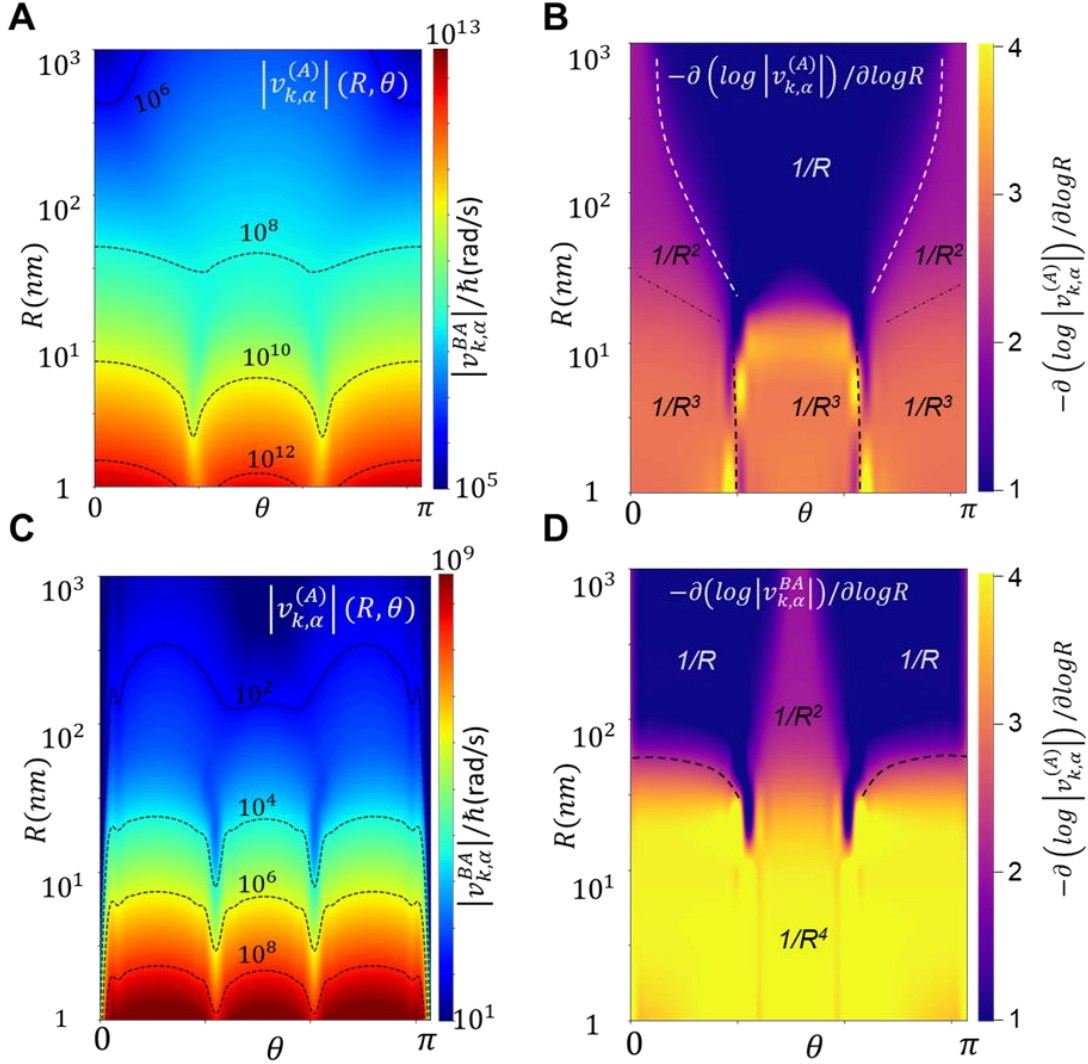

**Fig. 6.** The distribution of photon absorption matrix element $\left|v_{k,\alpha}^{(A)}(\bar{R})\right|$ for **(A)** spin conserving $|GS\rangle$ to $|{}^1T_{1u}\rangle$ and **(C)** spin non conserving $|GS\rangle$ to $|{}^3T_{1u}(m_s = \pm 1)\rangle$ transitions showing the response as a function of source-absorber distance R, and the polar angle $\theta$ (For these cases, cylindrical symmetry (Fig. 5) ensures no variation with $\phi$). Panel **(B)** and **(D)** show the plots of exponent $n$ of the $\sim 1/R^n$ dependence of $\left|v_{k,\alpha}^{(A)}\right|$ ($n = -\partial\left(\log\left|v_{k,\alpha}^{(A)}\right|\right)/\partial \log R$). The dashed lines are guide to the eye to indicate the boundary between far and near field regimes where the exponent $n$ changes its value.

We further explore $v_{k,\alpha}^{(A)}(R,\theta)$ for ED and MD transitions in Fig. 6. Figure 6(A) shows a 2D plot for $v_{k,\alpha}^{(A)}(R,\theta)$ for the $|GS\rangle$ to $|{}^1T_{1u}\rangle$ transition by an electric dipolar mode ($\alpha = ED$) as a function of the distance R and the polar angle $\theta$ of the source at which the absorber sits. In Figure 6(B) we plot $-\partial\left(\log\left|v_{k,\alpha}^{(A)}\right|\right)/\partial \log R$ for the same transition, which results in the dominant exponent $n$ corresponding to the $\left|v_{k,\alpha}^{(A)}\right|\sim 1/R^n$ relation. The dashed lines are guides to the eye. At the polar direction ($\theta = 0$ $and$ $\pi, \pm Z$ direction), the matrix element becomes zero at first order and thus the radial dependence is ~$1/R^2$ even at far field. In contrast, near the equator, ($\theta = \frac{\pi}{2}$, XY plane), a standard near field to far field transition of ~$1/R^3$ to ~$1/R$ behavior is observed for $\left|v_{k,\alpha}^{(A)}\right|$.

In Fig. 6(C) and 6(D), we show the results for a MD source inducing a spin non conserving $|GS\rangle$ to $|\,^3T_{1u}(m_s = \pm 1)\rangle$ transition. These results are markedly different from those obtained in the ED case. Now, the equator ($\theta \sim \frac{\pi}{2}$) is the zone where to the first order the matrix element vanishes and hence the radial dependence, even at R>10 nm, is ~$1/R^2$.

The results shown in Figures 4-6 map out for the first time the distance and angle dependence for the near field matrix elements pertaining to energy transfer processes, and involving spin allowed and spin forbidden transitions for both electric and magnetic dipole like sources. Our findings indicate that the symmetry of the angular distribution of the matrix elements corresponding to spin conserving and non-conserving transitions are different; hence our calculations point at possible design rules for geometrical configurations more prone to yield long-lived transitions useful to create desired states of memories or quantum networks in solid hosts.

**Transfer rate calculations**

We now turn to examining the NRET transfer rate. We consider a physical source of unity oscillator strength- which is a realistic oscillator strength for deep levels such as rare earth centers [7]. For a generic source of oscillator strength $f_{osc}$ the emission matrix elements satisfy

$$\sum_\alpha \left|v_{k\alpha}^{(S)}\right|^2 = \left(\frac{\mu_0 \hbar^2 \omega_k^2 e^2}{3\pi^2 m_0}\right) f_{osc}. \tag{14}$$

Here $\mu_0$ represents the vacuum permeability. Equation (14) yields the emission matrix element of a generic source if the energy $\omega_k$ and the oscillator strength are known.

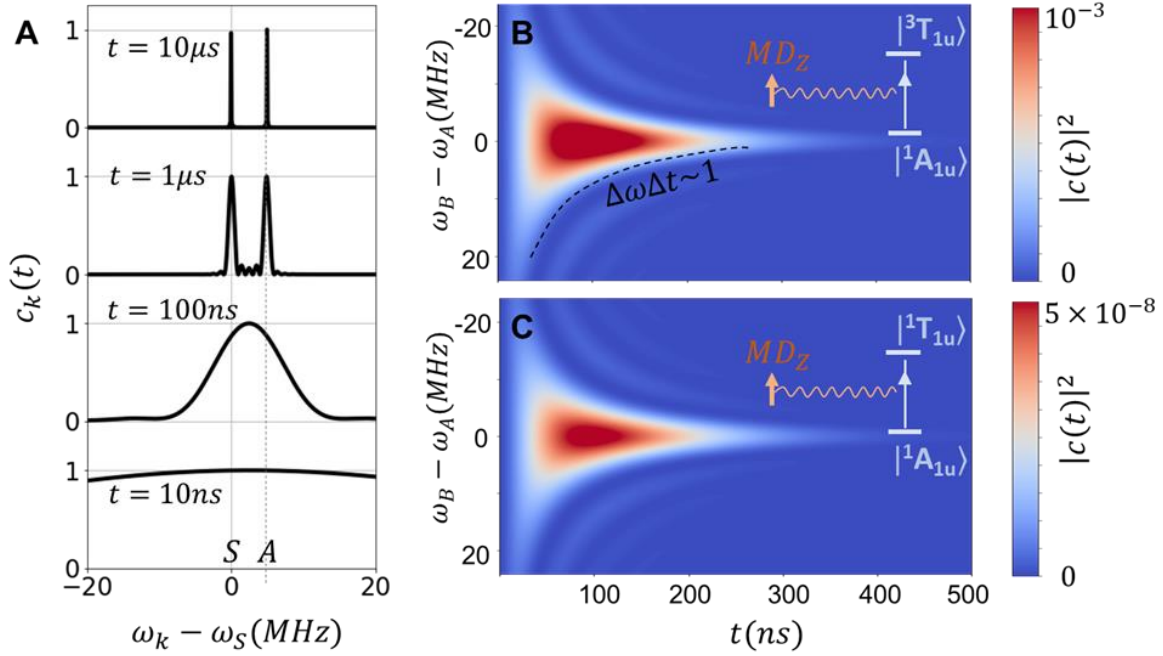

**Fig. 7. (A)** Normalized spectrum of the photon amplitude $c_k(t)$ as a function of the photon frequency ($\omega_k$) at different times = 10ns, 100ns, 1us, and 10us for the energy difference between source and absorber transition $\omega_A - \omega_S = 5 MHz$. **(B)** Two-dimensional plot of the NRET probability ($|c(t)|^2$) as a function of time and frequency mismatch $\omega_A - \omega_S$ for a magnetic dipole source (MD) of oscillator strength 1, which induces a singlet-to-triplet transition. The corresponding singlet-to-singlet transition for the same MD source is shown in **(C)** which is much weaker compared to the singlet-to-singlet transition at this configuration.

For specific type of a multipole source ($\alpha$ specified), Eq.(14) can be used to calculate the absorption matrix element values. Here we show the transfer amplitude for a purely electric dipole source with $\alpha = \{L, J_z, P = 1, 0, -1\}$, and a purely magnetic dipole source with $\alpha = \{L, J_z, P = 1, 0, 1\}$. We consider short timescale processes for which there can be finite transfer even for a finite energy mismatch. At small time, $\Delta\omega\, t \ll 1$, the width of the spectrum of $c_k(t)$ (Eq. (13)) is relatively large and it gradually decreases as $t$ increases. This is shown in Fig. 7(A) for the spin non conserving $|GS\rangle$ to $|\,^3T_{1u}(m_s = \pm 1)\rangle$ transition, induced by a magnetic dipole source of oscillator strength 1. The corresponding NRET probability of coherent transfer is shown in Fig. 7(B) as a function of both time in the horizontal axis and the energy mismatch between the source and the absorber in the vertical axis. As a comparison we also show in Fig. 7(C) the transfer probability for the spin-conserving singlet-singlet transition for the same magnetic dipole source of 1 oscillator strength. While the energy-time uncertainty behavior in Fig. 7(C) is the same as in panel 7(B), the transfer amplitude is negligible compared to that of the spin non-conserving transition. This is an example of the spin non conserving transition dominating over the spin conserving one for a realistic near-field process.

**Effective parameter to characterize coherent transfer**

Historically, the energy transfer between emitters has been characterized using the notion of Forster radius [1]. Such radius is defined as the distance between a source and an absorber at which the energy transfer rate between the source and the absorber matches the rate of radiative decay of an isolated source- resulting in a 50% probability of the transfer. In most systems investigated in the chemistry literature, the energy transfer of interest is between a large number of sources and absorbers. Thus, averaging over angular coordinates has been always implicit. However, in the solid state, using thin film growth, nanofabrication, and spatially selective doping techniques, some control can be achieved on the relative placement of the source and the absorber sites. Thus, the angular dependence cannot be assumed to be averaged out. In addition, due to the large phonon induced broadening in room temperature measurements in solution chemistry and biological systems, the emission and absorption spectra are always incoherently broad, resulting in a dominant incoherent energy transfer.

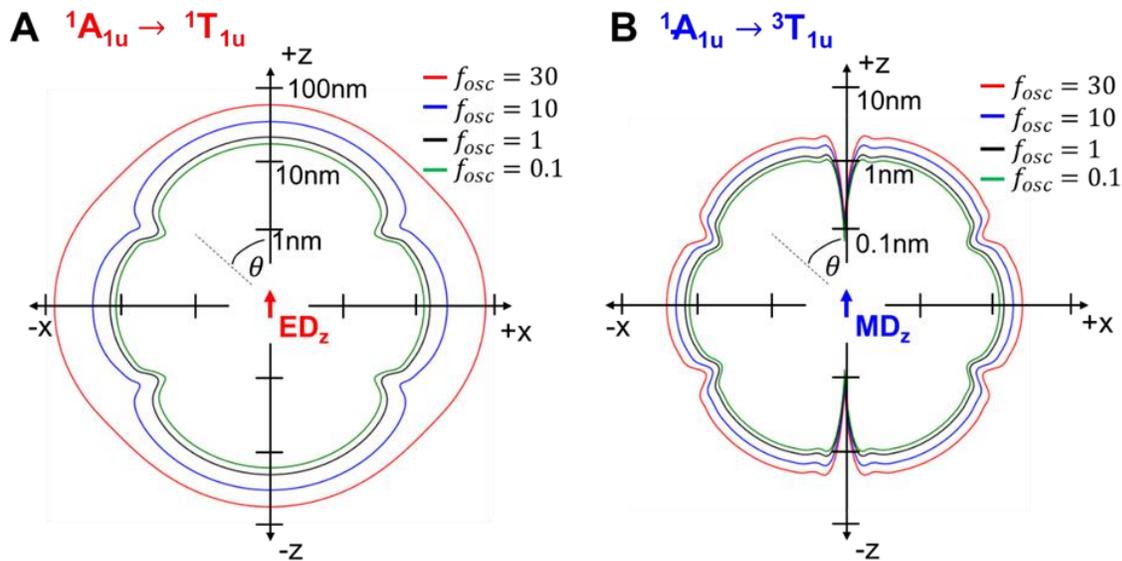

**Fig. 8.** Contour plots of the parameter $R_{eff}$ (see text) for the **(A)** singlet-singlet absorption transition by an F center from an electric dipole source, at near field, and the **(B)** singlet-to-triplet absorption from a magnetic dipole source. The different lines indicate cases with different oscillator strengths of the dipolar source: $f_{osc} \sim 0.1$ to $1$ represents a deep level source whereas, $f_{osc} \sim 10 - 30$ represents larger sources like quantum dots.

Our interest lies in a different regime. First, several deep level systems such as rare earth substitutional sites, NV centers, and other point defects, exhibit narrow emission spectra compared to the measurement timescale. Second, in the solid state most characterizations and device design occur at cryogenic temperature which suppresses decoherence and dephasing and can lead to even <kHz linewidths [6, 7]. Thus, in the following, we define a parameter representing the equivalent of a Forster radius for angle resolved configurations, but for coherent processes, and we include spin conserving and non-conserving transitions in the coherent limit.

We define the effective radius $R_{eff}$ as a function of the direction defined by the unit vector $\hat{R} = (\theta, \phi)$ from the source S to the absorber A such that $\max(|c(t)|^2) = 1/2$. Using equation (7) we get: $|c(t)|^2 = \frac{4M^2}{\hbar^2} \frac{\sin^2(\Delta\omega\, t/2)}{\Delta\omega^2} e^{-\frac{t}{T_{1S}}}$. The maximum probability is achieved at $t = t_{max} = 2\frac{\operatorname{atan}(\Delta\omega\, T_{1S})}{\Delta\omega}$. Imposing $|c(t)|^2 |_{t=t_{max}} = \frac{1}{2}$, and invoking $M = \frac{i\pi n_i}{\hbar c}\sum_\alpha v^{(S)}_{k_S,\alpha}\, v^{(A)}_{k_S,\alpha}$ we get:

$$\frac{\pi^2 n_i^2}{\hbar^2 c^2}\left|\sum_\alpha v^{(S)}_{k_A,\alpha}(R_{eff}, \hat{R})\, v^{(A)}_{k_A,\alpha}\right|^2 = \frac{\hbar^2(1+T_{1S}^2\Delta\omega^2)}{8T_{1S}^2}\, e^{\frac{t_{max}}{T_{1S}}} \qquad (15)$$

Solving equation (15) numerically for an arbitrary source and arbitrary absorber yields $R_{eff}$ as a function of the source-to-absorber direction $\hat{R}$.

We show $R_{eff}$ at resonant conditions ($\Delta\omega = 0$) for (1) the singlet ($|GS\rangle$) to singlet ($|\,^1T_{1u}\rangle$) absorption for a near field electric dipole source, and (2) the singlet ($|GS\rangle$) to triplet ($|\,^3T_{1u}(m_s = \pm 1)\rangle$) absorption for a near field magnetic dipole souce in Figure 8(A) and (B) respectively. The function $R_{eff}(\theta)$ is shown as a polar plot in the XZ plane—but a rotational symmetry around the Z axis can be assumed. The different lines in Fig. 8(A) and (B) represent different values of the oscillator strength of the source dipole, encompassing a vast category of emitters including deep levels ($f_{osc} \sim 0.1$ to $1$) and quantum dots ($f_{osc} \sim 10\ to\ 30$). We find that for the spin conserving

transition from an ED source of $f_{osc} = 1$, $R_{eff} \approx 10$ nm whereas for the spin non-conserving transition from a MD source, $R_{eff} \approx 2\ nm$. Distances of ~2 to 10 nm are realistic distances in either random ensembles or controlled pairs of defects and emitters in semiconductors and insulators. Additionally, with increasing $\Delta\omega$, as per Eq. (15) the variation of the matrix element follows approximately a Lorentzian and the $R_{eff}$ drops accordingly. We see from Fig.8 that in several cases, for both ED and MD sources, an angular coarse-graining may not be appropriate as there is a high degree of variation of $R_{eff}$ as a function of the source-absorber polar angle $\theta$. These results provide a possible way to understand quantitatively how to guide the growth and nano structuring of a solid host to facilitate specific, desired transitions.

## Discussion

We investigated the transfer of energy between localized emitters in a solid and explored the spin-non conserving transitions in the near field. The perturbative approach presented in this paper includes first-principle calculations of electronic states and accounts for the quantum nature of the broad spectrum virtual photons mediating the energy transfer. Our findings indicate that for realistic oxides, magnetic and spin non-conserving transitions are key to understand light-matter interaction and to derive design rules to engineer long-lived transitions. The approach provides a way to address a wide variety of NRET processes that are highly relevant to quantum and classical optical devices—specifically optical memories, quantum optical networks, and quantum memories.

Specifically, we applied our approach to the F center in MgO, and we presented a systematic study of the NRET amplitude for spin-conserving (singlet-to-singlet) and spin-non conserving (singlet-to-triplet) absorption transitions originating from a near field source (e.g. a RE impurity) of magnetic or electric dipole type, as a function of the relative orientation and distance between the source and the absorber. We showed that in certain configurations of the source and the absorber, singlet-to-triplet transitions not only become active at near field but constitute the dominant process. Our study revealed some key design principles to realize such singlet-to-triplet type transitions at the near field. In the case of the F center in MgO we found that a magnetic dipole type source is necessary to obtain dominant spin-non conserving transitions. In the case of rare earth doped oxides, an intrinsic magnetic dipole type emission is available [6,7], thus providing a viable path to create long-lived excited state in localized defects that are relevant to rare earth based ultra-high density optical memory platforms envisioned in this work [see SI]. Our approach can be used to facilitate material search to identify spectrally matched rare-earth-defect combinations for such optical memory platforms.

We also explored the NRET amplitude as a function of the angle between the source and the absorber and shown a significant angular variation of both the amplitudes and the radial dependence of the absorption matrix elements of the transitions. This variation represents a significant result as it provides insight into nanofabrication design where certain processes may be favored by engineering specific geometrical configurations of the emitters. Finally, we found that the effective radius for singlet-to-triplet transition in VO: MgO is ~ 1-2 nm for coherent transfer rate of 50% of the radiative decay, which provide insight into the required density of the source and absorber site in the host material to enable such processes.

Importantly, although we showed results for and exemplar oxide, VO: MgO, the approach presented here is general and applicable to any localized defects in any solid that can be described using first principles calculations. In addition, our approach is easily generalizable to distributed systems of defects thus allowing for the study of near field energy transfer in larger emitters such as quantum dots, as well as defects and dopants with inhomogeneous distributed spectral lines. These include the interaction between different REs for quantum memories, the interaction between a localized defect acting as a quantum memory with another localized defect acting as a qubit, and energy trapping by localized electronic states.

Our approach is suited for addressing photon hopping in large (~$\mu$m) ensembles, taking into account finite energy uncertainty, and including spin flip and spin conserving processes alike. The computational cost of the method presented here is only limited by the computation of the matrix elements at the separate localized emitters and thus large-scale systems with emitter-emitter distances ranging from ~1 nm to ~1000 nm can be tackled with equal computational cost.

Furthermore, using first principle electronic structure calculation paves ways to include lattice relaxation between energy transfer processes in a chain of emitters. The excited states of many localized emitters of interest undergo significant relaxation compared to the ground state. The approach reported here provides a pathway to understanding such physics of collective emitters in realistic devices towards classical and quantum photonic applications.

34. N. Sheng, C. Vorwerk, M. Govoni, G. Galli, Green's Function Formulation of Quantum Defect Embedding Theory. *J. Chem. Theory Comput.* **18**, 3512–3522 (2022).

35. C. Vorwerk, G. Galli, Disentangling photoexcitation and photoluminescence processes in defective MgO. *Phys. Rev. Materials*. **7**, 033801 (2023).

36. S. Verma, A. Mitra, Y. Jin, S. Haldar, C. Vorwerk, M. R. Hermes, G. Galli, L. Gagliardi, Optical Properties of Neutral F Centers in Bulk MgO with Density Matrix Embedding. *J. Phys. Chem. Lett.* **14**, 7703–7710 (2023).

37. Y. Jin, V. W. Yu, M. Govoni, A. C. Xu, G. Galli, Excited state properties of point defects in semiconductors and insulators investigated with time-dependent density functional theory (2023), (available at http://arxiv.org/abs/2309.03513).

38. E. E. Jelley, Spectral Absorption and Fluorescence of Dyes in the Molecular State. *Nature*. **138**, 1009–1010 (1936).

39. L. Stryer, R. P. Haugland, Energy transfer: a spectroscopic ruler. *Proc. Natl. Acad. Sci. U.S.A.* **58**, 719–726 (1967).

40. D. Klose, A. Holla, C. Gmeiner, D. Nettels, I. Ritsch, N. Bross, M. Yulikov, F. H.-T. Allain, B. Schuler, G. Jeschke, Resolving distance variations by single-molecule FRET and EPR spectroscopy using rotamer libraries. *Biophysical Journal*. **120**, 4842–4858 (2021).

41. F.-F. Kong, X.-J. Tian, Y. Zhang, Y. Zhang, G. Chen, Y.-J. Yu, S.-H. Jing, H.-Y. Gao, Y. Luo, J.-L. Yang, Z.-C. Dong, J. G. Hou, Wavelike electronic energy transfer in donor–acceptor molecular systems through quantum coherence. *Nat. Nanotechnol.* **17**, 729–736 (2022).

42. D. A. Gálico, M. Murugesu, Controlling the Energy-Transfer Processes in a Nanosized Molecular Upconverter to Tap into Luminescence Thermometry Application. *Angew Chem Int Ed*. **61** (2022).

43. A. Shukla, G. Kaur, K. J. Babu, A. Kaur, D. K. Yadav, H. N. Ghosh, Defect-Interceded Cascading Energy Transfer and Underlying Charge Transfer in Europium-Doped $CsPbCl_3$ Nanocrystals. *J. Phys. Chem. Lett.* **13**, 83–90 (2022).

44. K. N. Avanaki, G. C. Schatz, Entangled Photon Resonance Energy Transfer in Arbitrary Media. J. Phys. Chem. Lett. 10, 3181–3188 (2019).
26

**Acknowledgments**

We thank Prof. Jorge Sofo, Yu Jin, and Dr. Christian Vorwerk for useful discussions. We acknowledge the computational resources of the National Energy Research Scientific Computing Center (NERSC), a DOE Office of Science User Facility supported by the Office of Science of the U.S. Department of Energy under Contract No. DE-AC02-05CH11231, and the computational resources of the University of Chicago Research Computing Center (RCC).

**Funding:** Work supported by the U.S. Department of Energy, Office of Science, for support of microelectronics research, under contract number DE-AC0206CH11357.

## Supplementary Information

**S1. Near Field Energy Transfer in Ultra-High Density Optical Memory**

In this section we discuss the key concept behind a potential platform of ultra-high density optical memories with rare earth (RE) emitters and defects in oxides, a platform where the near field energy transfer plays a critical role. The approach exploits 4f-4f transitions of rare earths that lead to very narrow linewidth (sub kHz) even in solid state, with the typical inhomogeneous spread of these transitions being of ~MHz to ~GHz range [6,7]. The large inhomogeneous spread relative to the narrow homogeneous linewidth provides a way to optically address a large number (≳1000) of RE atoms even within a diffraction limited volume and provides a potentially path to ultra-high density optical memories.

However, the lifetime of the optically excited state is still limited by the radiative decay lifetime of the RE emitters (typically ~10 ms). We propose that a possible route to enhance the lifetime is to transfer the excitation to a proximal defect. Further, by inducing transitions that are spin-non conserving, i.e., optically forbidden in the far-field, one can potentially create longer lifetimes.

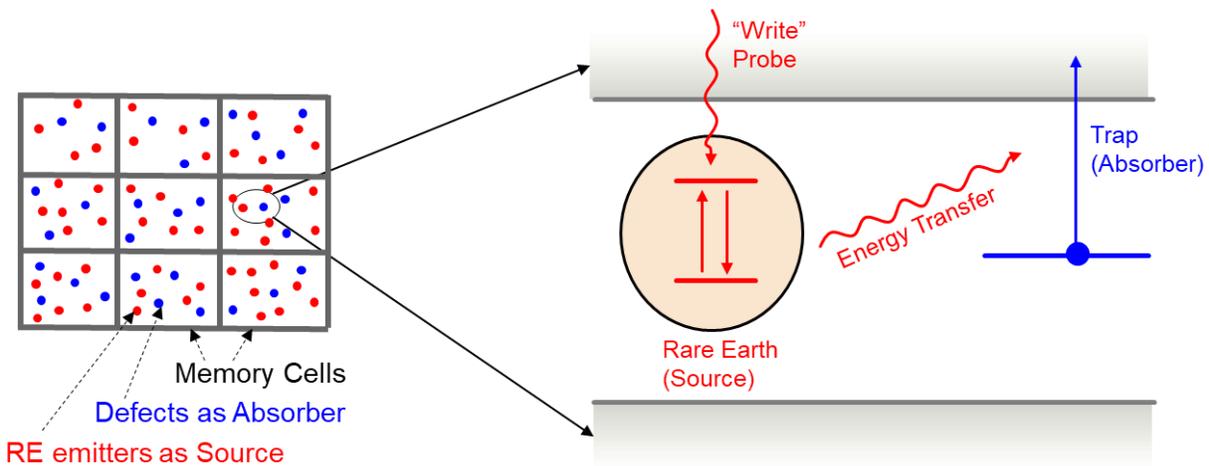

**Fig. S1.** A schematic representation of ultra-high density optical memories where each memory-cell in a solid host (left panel) contains an ensemble of RE emitters (red) and point-defects (blue). Right panel: A RE emitter- defect complex representing the working "unit" of the memory. Optical excitations of RE emitters that are spectrally separated can be transferred to a suitable defect in the proximity of the emitter to trap the excitation and increase its lifetime (see text).

The basic idea is captured in Fig. S1. Each cell in Fig. S1(A) represents an optically addressable memory cell doped with RE emitters (red) and suitable defects (blue) that exhibit optical absorption transitions in the same band at the RE emission. Individual RE ions can be identified by their narrow spectral lines and optically excited (Fig. S1(B)). The excitation can then be transferred and trapped into a nearby defect (e.g., an oxygen vacancy in an oxide). For typical doping concentration of ~few ppm, the average Res and vacancies can reach ~5 to ~10 nm- a distance much smaller than the wavelength of the optical/near IR photons (~500nm to ~1$\mu m$). Thus, understanding near field energy transfer is critical to understand the transfer process. Further, the excitation of the trap could be quenched, or Stark shifts could be exploited in the RE emission process as readout mechanisms.

Understanding the design rules required to engineer the processes described above call for (1) exploring possible combinations of RE emitters, defects, and host oxides to allow spectrally matched RE-defect combinations, and (2) a quantitative understanding of the near field energy transfer rates for optically allowed and optically forbidden absorption transitions in the near field and (3) understanding the energy trapping and relaxation dynamics. The combination of (1), (2), and (3) demands a unified approach between first principle electronic structure calculations and quantum electrodynamical near field energy transfer in the solid state- which is the framework established in this work.

**S2. Multipole Basis of the Photon**
We used a multipole basis to represent photon modes [54,55,56] in a uniform dielectric medium; this basis facilitates investigating the transfer/ conservation of angular momentum states during the energy transfer process [10] as all the multipole modes are eigenfunctions of the angular momentum operator. In this section we give a brief overview of how we represent photon modes.

In Figure S1 we consider a spherical cavity of radius $R_{norm}$ centered around the source denoted as S- here taken as a two-level system. The ladder-down operator of this 2-level system is $\hat{\sigma}_S$ and $\hat{a}^{(S)}{}_{k,\alpha}$ represents the annihilation operator of a photon mode with radial wavevector k and $\alpha = \{L, J_Z, P\}$ in the spherical cavity. Here $L$ denotes the orbital angular momentum of the photon, $J_Z$ the total (orbital + spin) angular momentum, and $P$ the parity. The multipole basis provides a complete orthogonal basis of the cavity photon mode and can be appropriately normalized for finite R as shown below. We then consider another two-level system, denoted as A, that acts as the absorber and has $\hat{\sigma}_A$ as the operator representing annihilation of an excitation.

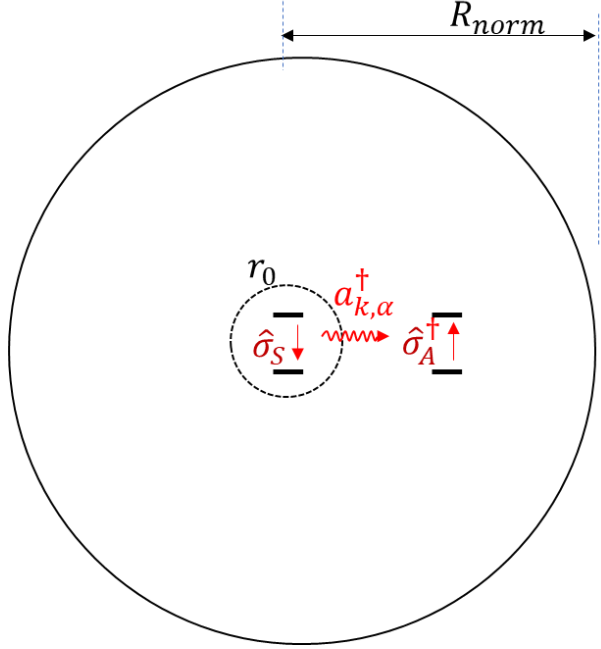

**Fig. S1**. Schematic representation of the macroscopic system considered in our work, where the photon modes are defined within a sphere of radius R centered at the emitter S. The absorber is denoted as A. The photon multipole modes are the propagating photon modes in a multipole basis (see text).

We express the overall Hamiltonian of the interacting source-absorber system (Eq. 2 of the main text) as:

$$H = H_S + H_A + H_{Field} + H_{int} \tag{S1}$$

In the 2$^{nd}$ quantized notation, we have:

$$\begin{aligned}
H &= \omega_S \hat{\sigma}_S^\dagger \hat{\sigma}_S + \omega_A \hat{\sigma}_A^\dagger \hat{\sigma}_A \\
&+ \sum_{k,\alpha} \omega_k \hat{a}^\dagger_{k,\alpha} \hat{a}_{k,\alpha} + \sum_{k,\alpha} (\hat{\sigma}_S + \hat{\sigma}_S^\dagger) \left( V_{k,\alpha}^{(S)} \hat{a}_{k,\alpha}^{(S)\dagger} + V_{k,\alpha}^{(S)\dagger} \hat{a}_{k,\alpha}^{(S)} \right) \\
&+ + \sum_{k,\alpha} (\hat{\sigma}_A + \hat{\sigma}_A^\dagger) \left( V_{k,\alpha}^{(A)} \hat{a}_{k,\alpha}^{(S)} + V_{k,\alpha}^{(A)\dagger} \hat{a}_{k,\alpha}^{(S)\dagger} \right)
\end{aligned} \tag{S2}$$

The multipole photon modes [55], i.e. $\left|1_{k,\alpha}^{(S)}\right\rangle = \hat{a}_{k,\alpha}^{(S)\dagger}|0\rangle$ are a complete basis that can be obtained by solving the Maxwell equations under the boundary condition chosen for a specific problem. The modes corresponding to $(-1)^L = P$ are of electric type and those corresponding to $(-1)^L = -P$ are referred to as magnetic-type multipoles [55].

We define the field operators for magnetic vector potential and scalar potentials, **A** and $A_0$ as $\bar{A}\left|1_{k,\alpha}\right\rangle = \bar{A}_{k,\alpha}(\bar{r},t)|0\rangle$ and $A_0\left|1_{k,\alpha}\right\rangle = A_{0,k,\alpha}(\bar{r},t)|0\rangle$. For a mode of electric type ($(-1)^L = P$) we have [55, 56]:

$$\bar{A}_{k,\alpha}(\bar{r},t) = \frac{1}{4\pi}\sqrt{\frac{k}{R_{norm}}}\left[\left(\sqrt{\frac{L}{2L+1}}\,g_{L+1}(kr)\,\bar{Y}_{L,L+1,J_z}(\hat{r}) + \sqrt{\frac{L+1}{2L+1}}\,g_{L-1}(kr)\,\bar{Y}_{L,L-1,J_z}(\hat{r})\right)\right.$$

$$+ C\left(-\sqrt{\frac{L+1}{2L+1}}\,g_{L+1}(kr)\,\bar{Y}_{L,L+1,J_z}(\hat{r})\right.$$

$$\left.\left.+ \sqrt{\frac{L}{2L+1}}\,g_{L-1}(kr)\,\bar{Y}_{L,L-1,J_z}(\hat{r})\right)\right]e^{-i\omega_k t} \qquad (S3)$$

And

$$\bar{A}_{0,k\alpha}(\bar{r}) = \frac{C}{4\pi}\sqrt{\frac{k}{R_{norm}}}\,g_L(kr)\,Y_{L,J_z}(\hat{r})e^{-i\omega_k t} \qquad (S4)$$

For a mode of magnetic type ($(-1)^L = -P$), we have:

$$\bar{A}_{k\alpha}(\bar{r}) = \frac{1}{4\pi}\sqrt{\frac{k}{R_{norm}}}\,g_L(kr)\,\bar{Y}_{L,L,J_z}(\hat{r})e^{-i\omega_k t} \qquad (S5)$$

and,

$$\bar{A}_{0,k\alpha}(\bar{r}) = 0 \qquad (S6)$$

where $g_L(kr) = 4\pi i^L z_L(kr)$, $z_L(kr)$ are spherical Bessel functions and $Y_{L,M}$ and $\bar{Y}_{J,L,M}$ are spherical scalar and vector harmonics. The constant C originates from the gauge freedom for the electric type multipoles and determines the amplitude of the scalar and the longitudinal photons [55]. $R_{norm}$ is the radius of the normalizing sphere for the multipole modes which translates to the width of the modes in k-space: $\Delta k = \frac{\pi}{R_{norm}}$. It is apparent from equation (S3) to (S6) that the vector and scalar potentials of the photons are normalized by $\sqrt{\Delta k}$. The normalization constant is included in $V_{k,\alpha}^{(S)}$ and $V_{k,\alpha}^{(A)}$. We thus define: $V_{k,\alpha}^{(S)} = v_\alpha^{(S)}(k)\sqrt{\Delta k}$, and $V_{k,\alpha}^{(A)} = v_\alpha^{(A)}(k)\sqrt{\Delta k}$. This definition facilitates the k-space summation over all virtual photons as shown below.

### *S2.1 Closed versus open systems*

As mentioned above, the photon modes $\left|1_{k,\alpha}^{(S)}\right\rangle$ are solutions of the Maxwell equation under properly chosen boundary conditions. In the case of perfectly reflecting boundary—e.g., a perfectly enclosing spherical mirror of radius $R_{norm}$, the radial functions are the Bessel function of type j- resulting in standing wave-type modes of a non-decaying photon. The Bessel j radial dependence can be decomposed into Hankel functions of type 1 and type 2- representing the radially inward propagating and outward propagating waves using the relation: $j_L(kr) = \frac{1}{2}\left(h_L^{(1)}(kr) + h_L^{(2)}(kr)\right)$ [55, 56].

If the boundary of the system is open, the radially outward propagating waves are represented by Hankel function of type 1. In this case, $g_L(kr) = 4\pi i^L h_L^{(1)}(kr)$. Such functions possess a singularity at the origin (phase center). However, this singularity does not affect the evaluation of the emission matrix elements. Thus, we can express the photon mode close to the center ($r < r_0$, for some $r_0$) by the Bessel j function. Hence for an open system, the photon mode $|1_{k,\alpha}\rangle$ represents the radially outward propagating wave, with type 1 Hankel function being the radial function in the source-free region ($r > r_0$), and the Bessel j type radial function near the source at $r < r_0$. [55].

From equation S2, we rewrite the interaction part of the Hamiltonian as:

$$H_{int} = \left(\sum_{k,\alpha} V_{k,\alpha}^{(S)} \hat{\sigma}_S \, \hat{a}^{(S)\dagger}{}_{k,\alpha} + \sum_{k,\alpha} V_{k,\alpha}^{(A)} \hat{\sigma}_A^\dagger \, \hat{a}_{k,\alpha}^{(S)}\right) + h.c.$$

$$+ \left(\sum_{k,\alpha} V_{k,\alpha}^{(S)} \hat{\sigma}_S \, \hat{a}^{(S)}{}_{k,\alpha} + \sum_{k,\alpha} V_{k,\alpha}^{(A)} \hat{\sigma}_A^\dagger \, \hat{a}_{k,\alpha}^{(S)\dagger}\right) + h.c. \quad (S7)$$

Here $V_{k,\alpha}^{(S)}$ and $V_{k,\alpha}^{(A)}$ are the matrix elements corresponding to the photon emission and photon absorption at S and A respectively that we evaluate in the main text.

### S3. Description of the Electron States

The source S and the absorber A may possess generic many-electron states; in general we have $H_X|\Phi^{(X)}\rangle = E^{(X)}|\Phi^{(X)}\rangle$ where $X = S, A$. The many electron eigenstates $|\Phi^{(X)}\rangle$ can be represented by a sum of slater determinants constructed from single electronic states $\{|\phi_i^{(X)}\rangle, i = 1:N_X\}$ where $N_X$ is the number of the single particle electronic states in emitter $X$. The Hamiltonians $H_S$ and $H_A$ may be approximated by the Kohn-Sham Hamiltonian in density functional theory [27-29], or with effective Hamiltonians derived from a chosen active space using, e.g. using quantum embedding theories [33-36]- depending on the level of electronic structure theory needed. The ground state of S or A can be generically represented as:

$$|GS^{(S/A)}\rangle = \sum_{\substack{i \in occ \\ j \in unocc}} \alpha_{ij,S/A}^{(GS)} c_j^{(S/A)\dagger} c_i^{(S/A)} |D^{(S/A)}\rangle \tag{S8}$$

and similarly, the excited state

$$|ES^{(S/A)}\rangle = \sum_{\substack{i \in occ \\ j \in unocc}} \alpha_{ij,S/A}^{(ES)} c_j^{(S/A)\dagger} c_i^{(S/A)} |D^{(S/A)}\rangle \tag{S9}$$

where $|D\rangle$ represents the Slater determinant built from the first filled N orbitals, i.e., $|D^{(S/A)}\rangle = \prod_{i=1}^{N} c_i^{(S/A)\dagger} |0\rangle$. Here $c_i^{(S/A)}$ denotes the annihilation operator of an electron in the single electronic state $|\phi_i^{(S/A)}\rangle$.

### S4. Matrix Elements for optically allowed and forbidden transitions of V$_O$: MgO

The matrix elements for transitions from a many-body ground to an excited state can be obtained by adding up all possible single-orbital matrix elements based on the Slater-Condon rule. For a transition from any generic many-body state $|\Phi\rangle$ to the state $|\Phi_1\rangle = c_j^\dagger c_i |\Phi\rangle$, we have:

$$\langle \Phi_1|\hat{O}|\Phi\rangle = \langle \phi_j|\hat{O}|\phi_i\rangle = O_{ji} \tag{S10}$$

Thus, for a generic single excited state: $|ES^{(A)}\rangle = \sum_{\substack{i \in occ \\ j \in unocc}} \alpha_{ij} c_j^{(A)\dagger} c_i^{(A)} |GS^{(A)}\rangle$, we have:

$$\langle ES^{(A)}|\widetilde{H}_{int}|GS^{(A)}, 1_{k,\alpha}^{(A)}\rangle = \sum_{\substack{i\in occ \\ j\in unocc}} \alpha^*{}_{ij}\langle \phi_j^{(A)}|\widetilde{H}_{int}| \phi_i^{(A)}, 1_{k,\alpha}^{(A)}\rangle \tag{S11}$$

For the $V_O$: MgO center, we use equation (S11) to write the matrix elements for the optical absorption singlet-to-singlet and singlet-to-triplet transitions below. The electronic defect states are localized mid-gap s-type orbital ($|s\rangle$) and localized p-type orbitals ($|p_x\rangle, |p_y\rangle, |p_z\rangle$) just above the conduction band edge. In the ground state configuration of the neutral F center, both spin states of the s-orbitals are filled resulting in a singlet ground state $|GS\rangle = |s_\uparrow, s_\downarrow\rangle$. For a excitation to the $|p_z\rangle$ orbital, the excited singlet can be written as $|\,^1T_{1u}\rangle = \frac{1}{\sqrt{2}}(|p_\uparrow, s_\downarrow\rangle + |s_\uparrow, p_\downarrow\rangle)$ whereas the three triplet states are: $|\,^3T_{1u,m_s=0}\rangle = \frac{1}{\sqrt{2}}(|p_\uparrow, s_\downarrow\rangle - |s_\uparrow, p_\downarrow\rangle)$, $|\,^3T_{1u,\,m_s=1}\rangle = |s_\uparrow, p_\uparrow\rangle$, and $|\,^3T_{1u,\,m_s=-1}\rangle = |p_\downarrow, s_\downarrow\rangle$. The matrix elements between these Slater determinants can be reduced to the matrix elements between the single electron orbitals in the following way:

$$\langle\,^1T_{1u}|H_{int}|GS, 1_{kLJzP}\rangle = \frac{1}{\sqrt{2}}\langle p_\uparrow|H_{int}|s_\uparrow, 1_{kLJzP}\rangle + \frac{1}{\sqrt{2}}\langle p_\downarrow|H_{int}|s_\downarrow, 1_{kLJzP}\rangle \tag{S12}$$

$$\langle\,^3T_{1u,m_s=1}|H_{int}|GS, 1_{kLJzP}\rangle = \langle p_\uparrow|H_{int}|s_\downarrow, 1_{kLJzP}\rangle \tag{S13}$$

$$\langle\,^3T_{1u,\,m_s=-1}|H_{int}|GS, 1_{kLJzP}\rangle = \langle p_\downarrow|H_{int}|s_\uparrow, 1_{kLJzP}\rangle \tag{S14}$$

$$\langle\,^3T_{1u,\,m_s=0}|H_{int}|GS, 1_{kLJzP}\rangle = \frac{1}{\sqrt{2}}\langle p_\uparrow|H_{int}|s_\uparrow, 1_{kLJzP}\rangle - \frac{1}{\sqrt{2}}\langle p_\downarrow|H_{int}|s_\downarrow, 1_{kLJzP}\rangle \tag{S15}$$

Note that the $\pm 1$ spin triplet state transitions are spin forbidden at far field, but they can have nonzero contributions at near field. The $m_s = 0$ triplet state transition vanishes if the symmetry between the up and down spin states is not broken. However, for a photon state with preferred direction of rotation along z, the symmetry between the up and down spin is broken and the matrix element for the $S_z = 0$ triplet state (Eqn. (S15)) also becomes nonzero.

**S5. Absorption Matrix Element for a Model System: The Simple Harmonic Oscillator**

For comparison with the results presented in the main text, in this section we show the absorption matrix elements of a model system with single-particle localized eigenstates. We choose a ubiquitous reference system i.e., the quantum simple harmonic oscillator (SHO). The SHO is defined by the Hamiltonian $H = \frac{p^2}{2m_e} + \frac{1}{2}m_e\omega^2 r^2$ where the energy eigenvalues are given by $E_{nx,ny,nz} = \hbar\omega\left(n_x + n_y + n_z + \frac{1}{2}\right)$, $n_{x/y/z}$ being the quantization number along X, Y and Z, and the eigenfunctions are given by the Hermite polynomials $\mathcal{H}$, i.e. $\langle r|\phi_{nx,ny,nz}\rangle \sim \mathcal{H}_{n_x}(x)\mathcal{H}_{n_y}(y)\mathcal{H}_{n_z}(z)$ with appropriate normalization. The energy spacing between the ground state $|\phi_{0,0,0}\rangle$ and the z-polarized first excited state $|\phi_{0,0,1}\rangle$ is $\hbar\omega$ which we equate to $5eV$ to match the absorption line of the F center in MgO. We then compute the matrix element between $|\phi_{0,0,0}\rangle$ and $|\phi_{0,0,1}\rangle$: $v_{k\alpha}^A = \langle\phi_{0,0,1}|H_{int}|\phi_{0,0,0}\rangle/\sqrt{\Delta k}$. Such a transition has, by definition, an oscillator strength of 1.

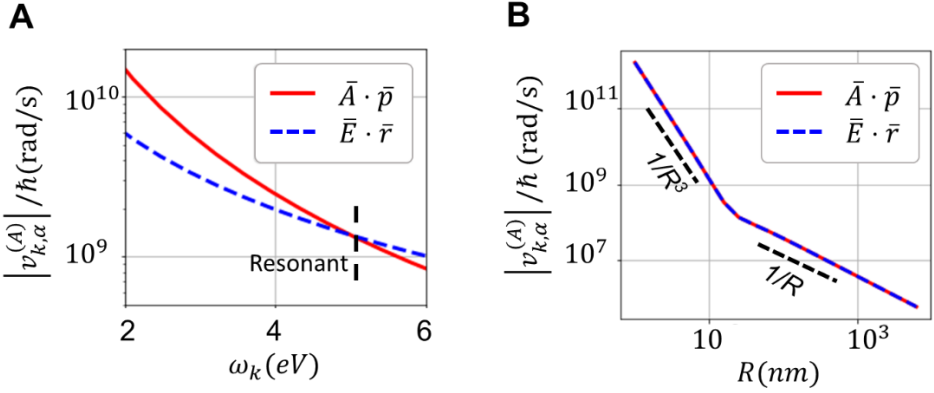

**Fig. S2.** Comparison between the absorption matrix element ($v_{k\alpha}^{BA}$) computed using the $A \cdot p$ Hamiltonian (red curve) and $E \cdot r$ Hamiltonian (blue curve) for the $|\phi_{0,0,0}\rangle$ to $|\phi_{0,0,1}\rangle$ transition at 5eV in an harmonic oscillator. Panel (A) shows the frequency response at a constant source-to-absorber distance of 10 nm and panel (B) shows the distance dependence at resonance (5eV).

As expected, in Fig. S2(A) it is apparent that the $\bar{A} \cdot \bar{p}$ and $e\bar{E} \cdot \bar{r}$ matrix elements are only equivalent at resonance. For the simple harmonic oscillator, the electron Hamiltonian is of the type of $H_A = \frac{p^2}{2m_e} + V(r)$ – hence the commutation relation $[H, \bar{r}] = \bar{p}$ holds which makes the two matrix elements equal at resonance ($\omega_k = \omega_A$). We further note from Fig. S2(B) that the equality between the $\bar{A} \cdot \bar{p}$ and $e\bar{E} \cdot \bar{r}$ matrix elements at resonance extend at both the near field and far field regimes. This is unlike the matrix element between the Kohn-Sham orbitals of the $V_O$: MgO defect shown in Fig. 2 of the main text, for which the Hamiltonian is nonlocal and thus the $\bar{A} \cdot \bar{p}$ and $e\bar{E} \cdot \bar{r}$ matrix elements are not equivalent.

## S6. Density Functional Theory Calculation of the Orbitals

We estimated the wavefunctions of the localized s- and p-orbitals of the $V_O$: MgO (as shown in Fig. 2) using Kohn-Sham Density Functional theory using the Quantum Espresso package [29]. We used the SG-15 norm-conserving Vanderbilt pseudopotentials [74]. We used both PBE [75] and dielectric dependent hybrid (DDH)[76] exchange correlation functionals and we found that this choice does not significantly affect the shape and localization of the defect orbitals- as indicated in Fig. 2 of the main text. There are several many body perturbation theory approaches to estimate the absorption and emission energies of the transitions [35, 36, 37]. However, in our calculations we simply take the known experimental value of 5eV as the absorption energy in the transition of interest.

## S7. Solution of Energy Transfer under 2$^{nd}$ Order Perturbation

We start with emitter S in the excited state at time t = 0 and derive the probability amplitude for the energy transfer to the absorber A as a function of time within 2$^{nd}$ order perturbation theory.

We consider two possible paths (Fig S3A). In path 1, the source S emits a photon and transitions from an excited state to the ground state and the photon is absorbed in A. This is shown in the Feynman diagram in Fig. S3(B). The interaction terms for this path come from the first term of the interaction Hamiltonian in Eq. (S7). In path 2, the time ordering of the photon absorption and emission events are reversed. In our case path 2 does not contribute to any NRET amplitude and only the contribution of path 1 provides the transfer amplitude [4, 5].

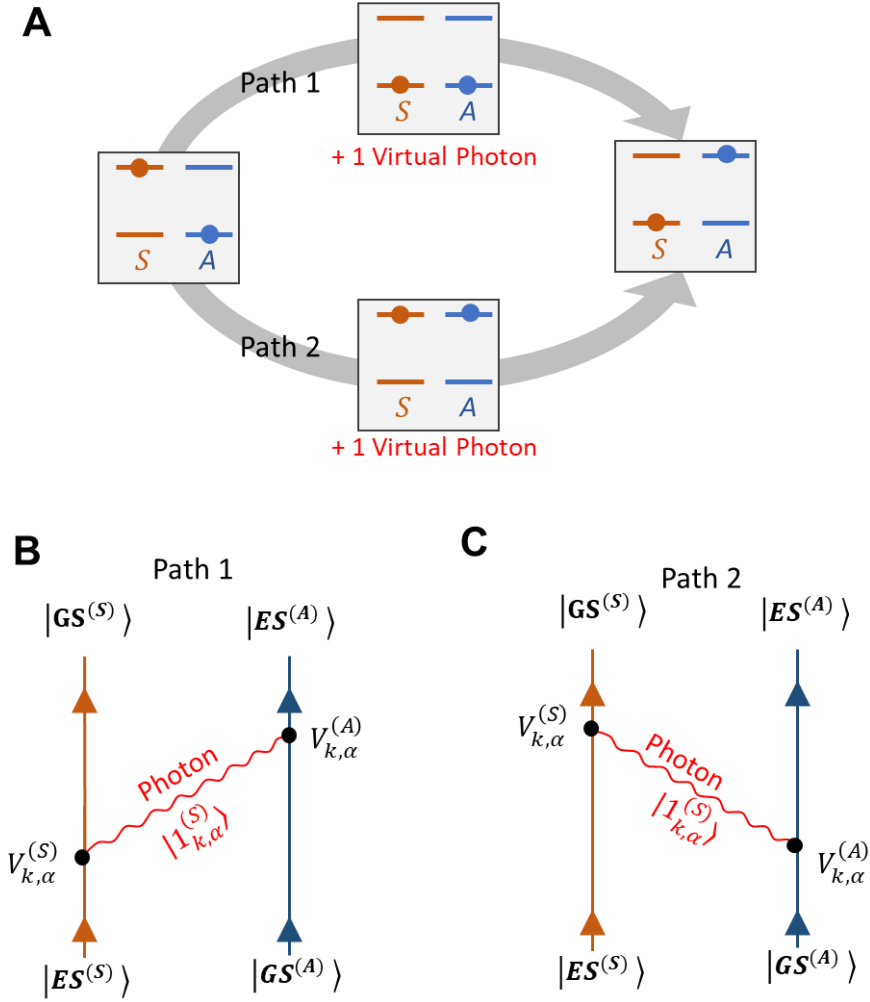

**Fig. S3.** (A) Two possible paths for the energy transfer from S to A, with respective Feynman diagrams shown in (B) and (C).

The overall NRET amplitude $c(t)$ can be expressed as:

$$c(t) = \langle \Psi_f | T\exp\left(-\frac{i}{\hbar} \int_0^t dt'\, \widetilde{H}_{int}(t')\right) | \Psi_i \rangle \tag{S16}$$

Here $\widetilde{H}_{int}(t') = e^{\frac{iH_0 t}{\hbar}} H_{int}(t') e^{-\frac{iH_0 t}{\hbar}}$ where $H_{int}(t')$ is from Eq. 1 and $H_0 = H_A + H_B + H_{Field}$ is the Hamiltonian of the unperturbed system. The initial state is $|\Psi_i\rangle = \Psi(t=0) = |ES^{(S)}\rangle|GS^{(A)}\rangle$; and the final state is $|\Psi_f\rangle = |GS^{(S)}\rangle|ES^{(A)}\rangle$. $T\exp$ represents exponential with time-ordering of the operators. Using 2$^{nd}$ order perturbation theory, we obtain:

$$c^{(2)}(t) = e^{-i\omega_B t}\left(-\frac{i}{\hbar}\right)^2 \int_0^t dt' \int_0^{t'} dt'' \sum_I \langle \Psi_f | \widetilde{H}_{int}(t') | I \rangle \langle I | \widetilde{H}_{int}(t') | \Psi_i \rangle \tag{S17}$$

The sum is over all possible intermediate states, which, for the positive time solution (path1, Fig. S3(A)) is given by $|I\rangle = |GS^{(S)}\rangle|GS^{(A)}\rangle|1_{k_A,\alpha}^{(S)}\rangle$. Thus, the evaluation of the NRET amplitude rests on the evaluation of the photon emission and absorption matrix elements denoted by $\langle I | \widetilde{H}_{int}(t') | \Psi_i \rangle$ and $\langle \Psi_f | \widetilde{H}_{int}(t') | I \rangle$ above. We calculate these matrix elements using wavefunctions obtained from first principle electronic structure theory based on DFT, as discussed in the main text. We have:

$$c^{(2)}(t) = -e^{-i\omega_A t}\frac{1}{\hbar^2}\sum_{k,\alpha} V_{k,\alpha}^{(S)} V_{k,\alpha}^{(A)} \int_0^t dt' \int_0^{t'} dt'' \, e^{-i(\omega_k-\omega_A)t'+i(\omega_k-\omega_S)t''}$$

$$= -e^{-i\omega_A t}\frac{1}{\hbar^2}\sum_{k,\alpha} \Delta k \, v_{k,\alpha}^{(S)} v_{k,\alpha}^{(A)} \int_0^t dt' \int_0^{t'} dt'' \, e^{-i(\omega_k-\omega_A)t'+i(\omega_k-\omega_S)t''} \tag{S18}$$

Note that the summation extends to all energies of the photon modes as well as all possible multipole modes of the photon. The summation in k can be converted to an integral in energies:

$$c^{(2)}(t) = \sum_k \Delta k \, c_k(t) = \int_0^\infty dk \, c_k(t) = \frac{n_i}{c}\int_0^\infty d\omega \, c_k(t) \tag{S19}$$

Here $c_k(t)$ is the contribution of the photons of wavevector k and width $\Delta k$ in k-space.

Straightforward evaluation of Eq.(S19) yields:

$$c_k(t) = \frac{1}{\hbar^2}\sum_\alpha \frac{v_{k\alpha}^{(S)} v_{k\alpha}^{(A)}}{\omega_k - \omega_S}\left(\frac{e^{-i\omega_S t} - e^{-i\omega_A t}}{\omega_A - \omega_S} - \frac{e^{-i\omega_k t} - e^{-i\omega_A t}}{\omega_A - \omega_k}\right) \tag{S20}$$

Assuming a smooth function for $v_{k\alpha}^{(S)}$ and $v_{k\alpha}^{(A)}$, the k integral is done by using a contour integration approach resulting in

$$c(t) = \frac{2i\pi n_i}{\hbar c}\sum_\alpha v_{k_S\alpha}^{(S)} v_{k_S\alpha}^{(A)} \frac{\sin\left(\Delta\omega \frac{t}{2}\right)}{\Delta\omega} = \frac{2M \sin\left(\frac{\Delta\omega}{2} t\right)}{\hbar \Delta\omega} \tag{S21}$$

$M = \frac{i\pi n_i}{\hbar c} \sum_\alpha v^{(S)}_{k_S,\alpha} v^{(A)}_{k_S,\alpha}$ here is the NRET matrix element which is equivalent to the coupling energy between the source and the absorber due all the photon modes. For a point dipole source and absorber, M reduces to the dipole-dipole coupling energy of dipole moments $\bar{p}_1$ and $\bar{p}_2$ separated by a distance $\bar{R}$ and is given by:

$$M(\bar{R}) = \bar{p}_1 \cdot \bar{\bar{G}}(\bar{R}) \cdot \bar{p}_2 \qquad (S22)$$

with

$$\bar{\bar{G}}(\bar{R}) = \frac{I - 3\hat{R} \otimes \hat{R}}{4\pi\epsilon R^3} + \frac{k^2 e^{ikR}}{4\pi\epsilon R}\left[\left(1 + \frac{ikR - 1}{k^2 R^2}\right)I + \left(\frac{3 - 3ikR - k^2 R^2}{k^2 R^2}\right)\hat{R} \otimes \hat{R}\right] \qquad (S23)$$

being the electromagnetic Green dyadic for a homogeneous medium. Eq. (S22) to Eq. (S23) are commonly used in molecular QED [54] to calculate energy transfer within the dipole approximation. However, unlike the dipole-dipole coupling, equation (S21) accounts for all the multipole modes and uses the absorption and emission matrix elements derived from first-principle theory.